\documentstyle[preprint,aps,eqsecnum]{revtex}

\begin{document}
\draft
\title{Stochastic semiclassical equations for weakly inhomogeneous
       cosmologies}
\author{Antonio Campos\footnote{Also at Institut de F\'{\i}sica
                                        d'Altes Energies (IFAE)}}
\address{Grup de F\'{\i}sica Te\`orica, Universitat Aut\`onoma de
        Barcelona, 08193 Bellaterra (Barcelona), Spain}
\author{Enric Verdaguer\footnotemark[1]}
\address{Departament de F\'{\i}sica Fonamental, Universitat de
        Barcelona, Av. Diagonal 647, \mbox{08028 Barcelona}, Spain}
\maketitle
\begin{abstract}
\end{abstract}
\baselineskip=15pt
\begin{quote}
\noindent  Semiclassical Einstein-Langevin equations for arbitrary
small metric perturbations conformally coupled to a massless quantum
scalar field in a spatially flat cosmological background are
derived. Use is made of the fact that for this problem the in-in or
closed time path effective action is simply related to the Feynman
and Vernon influence functional which describes the effect of the
``environment'', the quantum field which is coarse grained here, on
the ``system'', the gravitational field which is the field of
interest. This leads to identify the dissipation and noise kernels
in the in-in effective action, and to derive a
fluctuation-dissipation relation. A
tensorial Gaussian stochastic source which couples to the Weyl
tensor of the spacetime metric is seen to modify the
usual semiclassical equations which can be viewed now as mean field
equations. As a simple application we derive the correlation
functions of the stochastic metric fluctuations produced in a flat
spacetime with small metric perturbations due to the quantum
fluctuations of the matter field coupled to these perturbations.
\end{quote}


\section{Introduction}
\label{sec:introduction}


The semiclassical equations of gravity are the generalization of
Einstein equations when the matter source is described quantically.
The source of these equations is the vacuum expectation value of the
stress tensor of the matter fields. This theory which assumes that
the gravitational field is classical should work well outside the
Planck scales and when the quantum fluctuations are small because
the matter source is an average term. But as it has been emphasized
by several authors \cite{WFK} one should expect that,
if the fluctuations above the average are large, and this depends
on the quantum state of the fields, the semiclassical equations
should not give a correct description. In fact, one expects that
a better approximation should describe the gravitational field
in a probabilistic way. In other words, the semiclassical equations
should be substituted by some Langevin type equation with a
stochastic source which describes classically the quantum
fluctuations. A significant step in this direction was made by
Hu \cite{H89} who proposed to view the semiclassical back-reaction
problem in the framework of quantum open systems, where the
quantum fields are seen as the ``environment'' and the gravitational
field is seen as the ``system''. Following this proposal a
systematic study of the connection between quantum field theory
and statistical mechanics has resulted in the development of a
consistent mathematical framework in which the relevant
semiclassical Einstein-Langevin equations can be derived
\cite{CH94}\cite{HS95}\cite{HM95}. The key to these results is the
influence functional method \cite{FVH} used in nonequilibrium
statistical mechanics to describe the environment-system
interaction when only the state of the system is of interest.
Here, the environment are the quantum
fields whose degrees of freedom are not of interest in the
semiclassical back-reaction problem and are traced out, {\it i.e.}
coarse grained, and the system is the gravitational
field which is the field of interest. The influence functional
provides information about the dissipation and the noise suffered by
the system, and the corresponding fluctuation-dissipation relation.
In this framework it is seen that external
stochastic sources with a given probability distribution
appear in the semiclassical equation for the gravitational field.
The origin of the noise acting on the system are the coarse
grained quantum fields whose degrees of freedom have been traced out.

In this paper we derive the semiclassical Einstein-Langevin
equations for arbitrary linear metric perturbations conformally coupled
to a massless quantum scalar field in a spatially flat cosmological
model. Semiclassical Einstein-Langevin equations have been recently
derived by Hu and Sinha \cite{HS95} for small anisotropies
conformally coupled to massless fields in a spatially homogeneous
background working in the framework of quantum cosmology, and by
Hu and Matactz \cite{HM95} who derived the semiclassical equations
for the scale factor in a spatially flat universe due to the
coupling of different quantum scalar fields.

We work in the ``in-in'' or CTP (closed time path) effective action
framework. The CTP effective action technique which was first
proposed by Schwinger \cite{SKC} is an effective action method
adapted to compute expectation values of quantum operators rather
than matrix elements as in the usual ``in-out'' effective action
method. It has been adapted by Jordan \cite{J86} and Calzetta and
Hu \cite{CH87} to a curved background and from it real
evolution equations which admit an initial value formulation can
be derived. It has been used to derive the semiclassical equations
for small anisotropies in an homogeneous background due to quantum
matter fields \cite{CH87}\cite{CH89} and it was also used in a
previous paper \cite{CV94}, referred from now on as Paper I, to
derive the semiclassical equations for arbitrary linear
perturbations in homogeneous backgrounds due to massless
conformally coupled matter fields. In this last case the CTP
effective action method leads in a very direct way to the
evaluation of the vacuum expectation value of the field stress tensor
which had been obtained previously by other methods \cite{HWS}.
The CTP effective action method is also currently used to study
dissipative effects produced by the inflaton oscillations around a
true vacuum \cite{P90}. These effects are of great
interest to understand the reheating mechanism in the inflationary
cosmological scenario.

The connection between the CTP effective action and the influence
functional was established by Calzetta and Hu \cite{CH94} in the
semiclassical context. It turns out that in this case there is a
very simple linear relation between the so called influence action
and the CTP effective action. That relation is not so simple if
one quantizes the gravitational field or one wants the field
equations for the mean quantum field on a given background
\cite{CH95} since then one has to advocate decoherence to go from
a quantum to a classical description. This, of course, is of great
interest for the study of structure formation in the early
universe. In our semiclassical case, however, the CTP effective
action already partially derived in Paper I gives all the necessary
information not only on the dissipation effects but also on the
noise due to quantum fluctuations. All we need to do is to identify
the corresponding dissipation and noise kernels.

The plan of the paper is the following. In Sec.
\ref{sec:CTP formalism} we briefly review the CTP functional
formalism and derive the complete CTP effective action
for arbitrary linear metric perturbations in a
homogeneous background. Note that in Paper I the CTP effective
action was only partially given since the complete action was not
necessary to derive the (mean field) semiclassical field equations.
In Sec. \ref{sec:influence action} the influence action
is derived, the dissipation and noise kernels are obtained and
the fluctuation-dissipation relation is given. In Sec.
\ref{sec:field equations} we derive the semiclassical
Einstein-Langevin equation and identify the stochastic source.
We show that a tensorial Gaussian stochastic source which couples
to the Weyl tensor of the spacetime metric is
the origin of noise in this case. Finally, in Sec.
\ref{sec:correlations} we apply the Einstein-Langevin equation
in a non cosmological background and derive the correlation
functions of the stochastic metric fluctuations produced in a flat
spacetime with small metric perturbations as a consequence of
the coupling of the quantum matter field to these perturbations.


\section{CTP effective action for weak inhomogeneities}
\label{sec:CTP formalism}


First, let us sketch the CTP functional formalism for the
evaluation of the CTP effective action; for a more detailed
exposition see Refs. \cite{J86}\cite{CH87}\cite{CV94}
\cite{P90(2)}. Let us consider the quantization of a scalar field
$\phi (x)$, the usual in-out effective action is based in the
generating functional $W[J]$ which is related to the vacuum
persistence amplitude in the presence of some classical source
$J(x)$ \cite{AL73}. From this functional one generates
matrix elements of the field $\phi (x)$ by functional derivation
with respect to $J(x)$.
Now in order to work with expectation values rather than matrix
elements one defines a new generating functional whose dynamics
is determined by two different external classical sources $J_+$
and $J_-$ by letting the in vacuum evolve independently under
these sources, {\it i.e.}
$
   {\rm exp}\{iW[J_+,J_-]\} =
        \sum_\alpha \langle 0,in|\alpha,T\rangle_{J_-}
                    \langle\alpha ,T|in,0\rangle_{J_+},
$
where we have assumed that $\{|\alpha,T\rangle\}$ is a complete
basis of eigenstates of the field operator $\phi (x)$ at some
future time $T$. This generating functional
admits a path integral representation which we may write in a compact
form as
\begin{equation}
   e^{iW[J_+,J_-]} =
        \int {\cal D}[\phi_+]{\cal D}[\phi_-]
             e^{i\left(S[\phi_+]-S[\phi_-]
                       +J_+\phi_+-J_-\phi_-\right)},
   \label{eq:generating functional}
\end{equation}
where $J_\pm\phi_\pm$ stands for $\int d^nxJ_\pm(x)\phi_\pm(x)$
and it is understood that we sum over all fields $\phi_+$ and
$\phi_-$ with negative and positive frequency modes, respectively,
in the remote past (these are the in boundary conditions) but
which coincide at time $t=T$ (in practice one usually takes
$T\rightarrow\infty$, the remote future).
Thus this path integral can be thought of as the path sum of two
different fields evolving in two different time branches: one going
forward in time in the presence of $J_+$ from the in vacuum to a
time $T$, and the other backward in time in the presence of $J_-$
from time $T$ to the in vacuum, with the  constraint that
$\phi_+(T,\mbox{\bf x})=\phi_-(T,\mbox{\bf x})$. For that reason
this formalism is called CTP functional formalism.
Note that we use an arbitrary number $n$ of spacetime dimensions
in order to perform dimensional regularization.

{}From this generating functional expectation values can be obtained.
Let us define
\begin{equation}
   \frac{\delta W[J_+,J_-]}{\delta J_\pm} \equiv
        \pm\bar{\phi}_\pm [J_+,J_-] ,
   \label{eq:inverse relation}
\end{equation}
and assume that these equations can be reversed, then
the CTP effective action is the Legendre transform of the
above functional
\begin{equation}
   \Gamma_{CTP}[\bar{\phi}_+,\bar{\phi}_-] =
        W[J_+,J_-] - J_+\bar{\phi}_+ +J_-\bar{\phi}_-  ,
   \label{eq:definition of the CTP effective action}
\end{equation}
where it is understood that the external sources are
functionals of the fields $\bar{\phi}_+$ and $\bar{\phi} _-$
through the definitions (\ref{eq:inverse relation}). From this
by functional derivation with respect to $\bar\phi_\pm$
we get the equations for the expectation values
$\bar\phi_\pm[J_+,J_-]$. The equation for
the vacuum expectation value of the field $\bar{\phi}[0]\equiv
\bar{\phi}_\pm[0,0]=\langle 0,in|\phi(x)|in,0\rangle $ is then
simply obtained imposing that $J_\pm =0$:
\begin{equation}
   \left.\frac{\delta\Gamma_{CTP}[\bar{\phi}_+,\bar{\phi}_-]}
              {\delta\bar{\phi}_+}
   \right|_{\bar{\phi}_\pm = \bar{\phi}[0]}
   = 0.
   \label{eq:definition of field equations bis}
\end{equation}

To simplify the notation it is useful at this stage to introduce a
more compact notation
${\cal S}[\phi_a]= S[\phi_+]-S[\phi_-]$,
$\phi_a(x)=\left( \phi_+, \phi_- \right)$, and
$J_a(x)   =\left( J_+, J_- \right)$,
where $a$ and $b$ take the two values $+$ and $-$, and introduce
the metric $c_{ab}=\mbox{\rm diag}(1,-1)=c^{ab}$ to lower and
raise $a$, $b$ indices.

Let us now proceed to the evaluation of the effective action
$\Gamma_{CTP}[\bar\phi_a]$ up to the one loop order,
which corresponds to the first order expansion of $W[J_a]$ in
powers of $\hbar$. As usual \cite{AL73} we solve
(\ref{eq:generating functional}) by the steepest descent method,
let $\phi^{(0)}_a(x)$ be the solutions of the classical field
equations, and expand the exponent in
(\ref{eq:generating functional}) about these background fields up
to the second derivative of ${\cal S}[\phi_a]$. The integration in
(\ref{eq:generating functional}) is now Gaussian and we can write
to this one loop order,
\begin{equation}
   e^{iW[J_a]}\simeq e^{iW^{(0)}[J_a]}
                     \left(\det A_{ab}(x,y)\right)^{-1/2}
   \label{eq:gaussian generating functional}
\end{equation}
where
$
   W^{(0)}[J_a]={\cal S}[\phi^{(0)}_a]+\int d^nx J^a\phi^{(0)}_a,
$
and $A_{ab}(x,y)$ is a $2\times 2$ matrix defined by
$A_{+-}(x,y)=A_{-+}(x,y)\equiv  0$, and
\begin{equation}
   A_{++}(x,y)\equiv
      \left.\frac{\delta^2S[\phi_+]}
                 {\delta\phi_+(x)\delta\phi_+(y)}
      \right|_{\phi_+=\phi^{(0)}_+}, \,\,
   A_{--}(x,y)\equiv -
      \left.\frac{\delta^2S[\phi_-]}
                 {\delta\phi_-(x)\delta\phi_-(y)}
      \right|_{\phi_-=\phi^{(0)}_-}.
   \label{eq:A matrix}
\end{equation}
In terms of the propagator $G(x,y)=A^{-1}(x,y)$ which is a
functional of the background fields $\phi^{(0)}_a(x)$ we can write
(\ref{eq:gaussian generating functional}) as
\begin{equation}
      W[J_a]\simeq W^{(0)}[J_a]- {i\over2}Tr(\ln G).
\end{equation}
The effective action which is a functional of $\bar\phi_a$, can now
be explicitly found to the same order. Using
(\ref{eq:inverse relation}),
(\ref{eq:definition of the CTP effective action}) and the fact that
$\bar\phi_a$ differs from $\phi^{(0)}_a$ by a term of order $\hbar$
one can show that
$
   W^{(0)}[J_a]\simeq {\cal S}[\bar\phi_a]+\int d^nx J^a\bar\phi_a,
$
and we have finally
\begin{equation}
   \Gamma_{CTP}[\bar\phi_a]\simeq
                           {\cal S}[\bar\phi_a]-{i\over2}Tr(\ln G).
   \label{eq:general CTP effective action}
\end{equation}

This formalism can be extended to curved spacetimes without
difficulties if the spacetime is globally hyperbolic \cite{J86}. The
hypersufaces of constant time are now Cauchy hypersurfaces, the
in and out states are defined in the Cauchy hypersurfaces
corresponding to the far past and far future respectively, and
the spacetime integral must be taken now with the correct volume
element.

Let us now compute explicity the CTP effective action for
a conformally coupled massless real scalar field in a nearly
conformally flat spacetime. Since the detailed calculations were
explained in Paper I, here we give only a summary of the main
results. Our spacetime background is a spatially flat FLRW universe
with small perturbations
\begin{equation}
   \tilde g_{\mu\nu}(x)=e^{2\omega(\eta)}
                                  \left(\eta_{\mu\nu}+h_{\mu\nu}(x)
                                  \right)
                       \equiv e^{2\omega(\eta)}g_{\mu\nu}
   \label{eq:FRWL}
\end{equation}
where $\eta_{\mu\nu}=\mbox{\rm diag}(-1,+1,\cdots,+1)$,
$\mbox{\rm exp}(2\omega(\eta))$ is the conformal factor,
$\eta$ is the conformal time which is related to the cosmological
time $t$ by $dt=\mbox{\rm exp}[\omega(\eta)]d\eta$, $h_{\mu\nu}(x)$
is a symmetric tensor which represents arbitrary small metric
perturbations and we have also introduced the nearly flat metric
$g_{\mu\nu}$ which is conformally related to $\tilde g_{\mu\nu}$.
The classical action for a free massless conformally coupled real
scalar field $\Phi(x)$ is given by
$S_m[\tilde g_{\mu\nu},\Phi]=
        -{1\over2}\int d^nx\sqrt{-\tilde g}
            \left[\tilde g^{\mu\nu}\partial_\mu\Phi\partial_\nu\Phi
                 +\xi(n)\tilde R\Phi^2
            \right],$
where $\xi(n)=(n-2)/[4(n-1)]$, and $\tilde R$ is the Ricci scalar
for the metric $\tilde g_{\mu\nu}$. Because of the conformal
coupling we can define a new field
$\phi(x)\equiv \mbox{\rm exp}[(n/2-1)\omega(\eta)]\Phi(x)$ and
the action $S_m$ after one integration by parts takes the form
\begin{equation}
   S_m[g_{\mu\nu},\phi] =
        -{1\over2}\int d^nx\sqrt{- g}
            \left[g^{\mu\nu}\partial_\mu\phi\partial_\nu\phi
                 +\xi(n) R\phi^2
            \right]
\end{equation}
which is the action for a free massless conformally coupled real
scalar
field $\phi(x)$ in a spacetime with metric $g_{\mu\nu}$, {\it i.e.}
a nearly flat spacetime in our case. Although the physical field
is $\Phi(x)$ the fact that it is related to the field $\phi(x)$
by a power of the conformal factor implies that a positive
frequency mode of the field $\phi(x)$ in flat spacetime will
correspond to a positive frequency mode in the conformally related
space. Thus non trivial quantum effects will be
due to the breaking of conformal flatness which in this case is
produced by the perturbations $h_{\mu\nu}(x)$. Let us expand the
above action in terms of these perturbations
\begin{equation}
   S_m[h_{\mu\nu},\phi]=\sum^\infty_{i=0}S^{(i)}_m[h_{\mu\nu},\phi]
   \label{eq:matter action}
\end{equation}
where the first term $(i=0)$ is simply the action for the field
$\phi$ in flat spacetime.

To the classical action for the matter fields $S_m$ we have to add
the action of the gravitational field $S_g[\tilde g_{\mu\nu}]$,
{\it i.e.} the Einstein-Hilbert action. In order to renormalize
the effective action later on we need to add appropriate terms
quadratic in the Riemann tensor (see Eq. (3.6) in Paper I). We
can also expand this action as
$
S_g[\tilde g_{\mu\nu}]\equiv
\sum^\infty_{i=0}S^{(i)}_g[\omega,h_{\mu\nu}]
$,
where we note that $S^{(0)}_g[\omega] $ depends on $\omega$ only.

We now introduce two scalar fields $\phi_+(x)$ and $\phi_-(x)$
which coincide at some future time $T$,
$\phi_+(T,{\bf x})=\phi_-(T,{\bf x})$, and which evolve in two
different geometries given by $h^+_{\mu\nu}$ and $h^-_{\mu\nu}$
such that $h^+_{\mu\nu}(T,{\bf x})=h^-_{\mu\nu}(T,{\bf x})$. The
CTP effective action for the gravitational and matter action can be
written as
\begin{equation}
   \Gamma_{CTP} = S_g [\omega ,h^+_{\mu\nu}]
                 -S_g [\omega ,h^-_{\mu\nu}]
                 +\Gamma_m [h^\pm_{\mu\nu},\bar\phi_\pm]
   \label{eq:CTP total}
\end{equation}
where $\Gamma_m $ contains the quantum effects of the scalar field
which one can compute using (\ref{eq:general CTP effective action}).

Note that since our action is quadratic in the fields the one loop
order result (\ref{eq:general CTP effective action}) is exact in this
case. Now the propagator $G_{ab}$ cannot be found exactly, thus we
expand $A_{ab}$ in the metric perturbations as
$
   A_{ab}= A^0_{ab} + V^{(1)}_{ab} + V^{(2)}_{ab}+...
$,
where $A^0_{ab}$ corresponds to the Minkowski background and
$V^{(i)}_{ab}$ are the inhomogeneous corrections, their explicit
values up to second order are given in Sec. III of Paper I.
Then we can write
$
   G_{ab}=G^0_{ac}\left[\delta_{cb}-V^{(1)}_{cd}G^0_{db}
                        -V^{(2)}_{cd}G^0_{db}
                        +V^{(1)}_{cd}G^0_{de}V^{(1)}_{ef}G^0_{fb}
                        +\cdots
                  \right]
$,
where the unperturbed propagator, defined by
$A^0_{ac}G^0_{cb}=\delta_{ab}$, is a matrix such that
$G^0_{++}=\Delta_F$, $G^0_{--}=-\Delta_D$, $G^0_{+-}=-\Delta^+$ and
$G^0_{-+}=\Delta^-$, where $\Delta_F$ and $\Delta_D$ are the
Feynman and Dyson propagators, respectively, and $\Delta^\pm$ are the
Wightman functions. Substituting this into
(\ref{eq:general CTP effective action}) and expanding its logarithmic
term we get up to second order in the metric perturbations,
\begin{eqnarray}
   \Gamma_m[h^\pm_{\mu\nu},\bar\phi_\pm]
        &\simeq &\sum^2_{i=0}\left(S^{(i)}_m[h^+_{\mu\nu},\bar\phi_+]
                                   -S^{(i)}_m[h^-_{\mu\nu},\bar\phi_-]
                             \right)
                 -{i\over2}Tr (\ln G^0_{ab})
            \nonumber \\
         & &+{i\over2}Tr
                    \left[\atop\right.
                         V^{(1)}_+G^0_{++}-V^{(1)}_-G^0_{--}
                        +V^{(2)}_+G^0_{++}-V^{(2)}_-G^0_{--}
            \nonumber \\
         & &\hskip1cm
                        -{1\over2}V^{(1)}_+G^0_{++}V^{(1)}_+G^0_{++}
                        -{1\over2}V^{(1)}_-G^0_{--}V^{(1)}_-G^0_{--}
                        +V^{(1)}_+G^0_{+-}V^{(1)}_-G^0_{-+}
                    \left.\atop\right],
         \label{eq:formal CTP effective action}
\end{eqnarray}
where we have defined $V^{(i)}_+\equiv V^{(i)}_{++}$ and
$V^{(i)}_-\equiv -V^{(i)}_{--}$. In Paper I we did not write the
terms which do not depend on the $\bar\phi_+$ field because such
terms were not needed to derive the field equation for the mean field
$\bar\phi[0]$, see (\ref{eq:definition of field equations bis}).

The explicit computation of the different terms was given in Paper I,
the new term
$-{i\over4}Tr (V^{(1)}_- \Delta_D V^{(1)}_- \Delta_D)$ can be
easily evaluated following closely that reference. After dimensional
regularization \cite{L75} of the divergent terms, renormalizing
with the action of the gravitational field and substituting the
field equations for $\bar\phi_{\pm}$ we get the renormalized
effective action for the gravitational field up to second order as
\begin{eqnarray}
   \Gamma ^{R}_{CTP}[ {\tilde g}^\pm_{\mu\nu}]
        &\equiv& S^{R}_{g}[{\tilde g}^+_{\mu\nu}]
           -S^{R}_{g}[{\tilde g}^-_{\mu\nu}]
           \nonumber \\
        & &+\frac{\alpha}{4}
           \int d^4xd^4y
               \left[3R^{+}_{\mu\nu\alpha\beta}(x)
                      R^{+\mu\nu\alpha\beta}(y)
                     -R^{+}(x)R^{+}(y)
               \right]\mbox{\rm K}^+(x-y;\bar\mu)
           \nonumber \\
        & &-\frac{\alpha}{4}
           \int d^4xd^4y
               \left[3R^{-}_{\mu\nu\alpha\beta}(x)
                      R^{-\mu\nu\alpha\beta}(y)
                     -R^{-}(x)R^{-}(y)
               \right]\mbox{\rm K}^-(x-y;\bar\mu)
           \nonumber \\
        & &+\frac{\alpha}{2}
           \int d^4xd^4y
               \left[3R^{+}_{\mu\nu\alpha\beta}(x)
                      R^{-\mu\nu\alpha\beta}(y)
                     -R^{+}(x)R^{-}(y)
               \right]\mbox{\rm K}(x-y),
   \label{eq:renormalized CTP}
\end{eqnarray}
where $\alpha = (2880\pi^2)^{-1}$, $\bar\mu$ is a renormalization
parameter,
\begin{eqnarray}
   \mbox{\rm K}^\pm(x-y;\bar\mu)
      &\equiv&
          -{1\over2}\int {d^4p\over(2\pi)^4} e^{ip\cdot(x-y)}
           \ln\left[ {(p^2\mp i\epsilon)\over\bar\mu^2}
              \right],
            \nonumber \\
   \mbox{\rm K}(x-y)
      &\equiv&
        -{1\over 2}\int {d^4p\over(2\pi)^4}
         e^{ip\cdot(x-y)}(2\pi i)\theta(-p^2)\theta(-p^0),
   \label{eq:K}
\end{eqnarray}
and $S^R_g[\tilde g^\pm_{\mu\nu}]$ are local terms coming from the
gravitational field only (see (3.29) in Paper I):
\begin{eqnarray}
   S^{R}_g[{\tilde g}_{\mu\nu}]
        &=&\int d^4x(-\tilde g(x))^{1/2}
              \left[ \frac{\tilde R(x)}{16\pi G_N}
                    -{\alpha\over 12} \tilde R^2(x)
              \right]
           \nonumber \\
        & &+2\alpha \int d^4x(-g(x))^{1/2}
              \left[ G^{\mu\nu}(x)\omega_{;\mu}\omega_{;\nu}
                    +\Box_g\omega(\omega_{;\nu}\omega^{;\nu})
                    +{1\over 2}(\omega_{;\mu} \omega^{;\mu})^2
              \right]
           \nonumber \\
        & &+\alpha\int d^4x(-g(x))^{1/2}
              \left[ R_{\mu\nu\alpha\beta}(x)
                     R^{\mu\nu\alpha\beta}(x)
                    -R_{\mu\nu}(x)R^{\mu\nu}(x)
              \right] \omega (x).
   \label{eq:renormalized gravitational action}
\end{eqnarray}
Here terms such as
$R_{\mu\nu\alpha\beta}$ refer to the metric $g_{\mu\nu}$ and are
linear in $h_{\mu\nu}$, and terms with tilde,
$\tilde R_{\mu\nu\alpha\beta}$, refer to the metric
$\tilde g_{\mu\nu}$.

It is remarkable that the $\Gamma_{CTP}$ is explicitly gauge
independent
since it is given in terms of the curvature tensor, {\it i.e.} it
is invariant under the infinitesimal coordinate change
$x^\mu \rightarrow x^\mu + \zeta^\mu(x)$, for arbitrary fields
$\zeta^\mu(x)$, which produce the change of $h_{\mu\nu}$ by
$h_{\mu\nu} + 2\zeta_{(\mu,\nu)}$ in the metric perturbations.
No gauge choice was made in the above calculation.


\section{Influence action for weak inhomogeneities}
\label{sec:influence action}


As explained in the introduction we now consider the interaction
of the scalar field $\phi(x)$ and the gravitational field from the
point of view of a quantum open system. The gravitational field
which is the field of interest to us will be the ``system'' and the
quantum scalar field will be the ``environment''. Since we wish to
know the effect of the environment on the system we will trace out the
degrees of freedom of the scalar field.

We will now summarize Calzetta and Hu \cite{CH94} arguments
leading to the relation between the CTP effective action
and the influence action in the semiclassical case.
We start by writting the generating
functional $W[J_+,J_-]$ for a system in which we have the
gravitational field, here symbolically called $g(x)$, and a scalar
field $\phi(x)$. We can write the classical action for these two
fields as: $S_g[g]+S_m[\phi]+S_{int}[g,\phi]$. Note that these terms
can be easily identified in our case from the expressions
of the matter and gravitational actions.

If we consider the field $\phi$ substituted by the two fields
$g(x)$ and $\phi(x)$ in (\ref{eq:generating functional}), we can
write
\begin{equation}
   e^{iW[J_+,J_-]} =
        \int {\cal D}[g^+]{\cal D}[g^-]
             e^{i\left(S_g[g^+]-S_g[g^-]
                       +J_+g^+-J_-g^-+S_{IF}[g^+,g^-,T]
                 \right)
               },
   \label{eq:generating functional gravity plus matter}
\end{equation}
where we have defined
\begin{equation}
   e^{i S_{IF}[g^+,g^-,T]} =
        \int {\cal D}[\phi_+]{\cal D}[\phi_-]
             e^{i\left(S_m[\phi_+]-S_m[\phi_-]
                       +S_{int}[g^+,\phi_+]-S_{int}[g^-,\phi_-]
                 \right)
               }.
   \label{eq:coarse grained}
\end{equation}
In (\ref{eq:coarse grained}) it is understood that we sum over all
fields $\phi_+$ and $\phi_-$ with negative and positive frequency
modes, respectively, in the remote past which coincide at time $T$
(remote future), and a similar interpretation is assumed in the path
integration of (\ref{eq:generating functional gravity plus matter}).
Since we are only interested in the gravitational field we couple
external sources $J_+(x)$, $J_-(x)$ to the fields $g(x)$ only
and not to the scalar fields $\phi(x)$. Note that we are using a
symbolic notation since the gravitational field is tensorial and
so are the external sources $J(x)$.

The interesting point here is that (\ref{eq:coarse grained}) under
the above interpretation is exactly the influence functional at
time $T$, ${\cal F}[g^+,g^-,T]$ as defined in \cite{FVH}, and thus
$S_{IF}[g^+,g^-,T]$ is the influence action. It is the action one has
to add to the classical actions $S_g[g^+]-S_g[g^-]$ to compute any
quantum probability for the transition from an initial state $g$ at
$t\rightarrow -\infty$ to a final state at the future time $T$.
It is also
the essential ingredient for the evolution operator, from initial
to final time, for the reduced density matrix in this case
\cite{CH94}. In this context the fields $+$ and $-$ and the sign
difference appear as a result of the double integration
needed to go from a transition amplitude to a probability.

We do not attempt to quantize the gravitational field even though
the formal path integration in
(\ref{eq:generating functional gravity plus matter}) seems to
indicate so. Such quantization is, of course, a highly non trivial
problem, we do not know the measure in
(\ref{eq:generating functional gravity plus matter}) and even in
the linear case we have to deal with the
gauge freedom. But if we restrict ourselves to the classical
approximation it is easy to find the CTP effective action
$\Gamma_{CTP}[g^+,g^-]$ from the generating functional
$W[J_+,J_-]$ in
(\ref{eq:generating functional gravity plus matter}) using the
Legendre transformation
(\ref{eq:definition of the CTP effective action}). We just need
to follow the steps which lead from
(\ref{eq:generating functional}) to
(\ref{eq:general CTP effective action}) but to zero order in
$\hbar$, thus we have simply
$\Gamma_{CTP}[g^+,g^-]=S_g[g^+]-S_g[g^-]+S_{IF}[g^+,g^-,T]$. On the
other hand the $\Gamma_{CTP}[g^+,g^-]$ in this equation is just the
renormalized CTP action $\Gamma^R_{CTP}[{\tilde g}^\pm]$ found in
(\ref{eq:renormalized CTP}) since that action follows from the
path integration (\ref{eq:coarse grained}) to one loop order for the
matter fields (including scalar sources) and the explicit
substitution of the field equations
for these quantum fields. The divergencies are removed by
appropriate terms in the classical gravitational action $S_g$ as
has been shown in Sec. \ref{sec:CTP formalism}; note that these
counterterms are implicitly assumed in
(\ref{eq:generating functional gravity plus matter}) to remove the
divergencies of (\ref{eq:coarse grained}). Thus we have \cite{CH94},
\begin{equation}
   \Gamma^R_{CTP}[{\tilde g}^\pm]
      =S^R_g[{\tilde g}^+]-S^R_g[{\tilde g}^-]+S^R_{IF}[h^+,h^-,T],
\end{equation}
where we have written $h^\pm$ in $S^R_{IF}$ to emphasize that it
depends on the metric perturbations only.
Comparing this expression with (\ref{eq:renormalized CTP}) we can
write the influence action as:
\begin{eqnarray}
   S^R_{IF}[h^{\pm}_{\mu\nu}]
        &=&\hskip.28cm\frac{3\alpha}{2}
           \int d^4xd^4y
               C^{+}_{\mu\nu\alpha\beta}(x)
               C^{+\mu\nu\alpha\beta}(y)
               \mbox{\rm K}^+(x-y;\bar\mu)
           \nonumber \\
        & &-\frac{3\alpha}{2}
           \int d^4xd^4y
               C^{-}_{\mu\nu\alpha\beta}(x)
               C^{-\mu\nu\alpha\beta}(y)
               \mbox{\rm K}^-(x-y;\bar\mu)
           \nonumber \\
        & &+3\alpha
           \int d^4xd^4y
               C^{+}_{\mu\nu\alpha\beta}(x)
               C^{-\mu\nu\alpha\beta}(y)
               \mbox{\rm K}(x-y).
\end{eqnarray}
Here we have introduced the Weyl tensor
$C_{\mu\nu\alpha\beta}(x)$, after using
the following expressions, which are easily shown to be satisfied
for an arbitrary function $f(x-y)$,
\begin{eqnarray}
   \int d^4xd^4y f(x-y)
        & &\left[  \,\,R_{\mu\nu\alpha\beta}(x)
                   R^{\mu\nu\alpha\beta}(y)
                 -4R_{\mu\nu}(x)
                   R^{\mu\nu}(y)
                 + R(x)
                   R(y)
         \right] = 0(h^3_{\mu\nu}),
        \nonumber \\
   \int d^4xd^4y f(x-y)
         &&\left[  \atop\right.
                   C_{\mu\nu\alpha\beta}(x)
                   C^{\mu\nu\alpha\beta}(y)
        \nonumber \\
         & &      -R_{\mu\nu\alpha\beta}(x)
                   R^{\mu\nu\alpha\beta}(y)
                 +2R_{\mu\nu}(x)
                   R^{\mu\nu}(y)
                 -\frac{1}{3}
                   R(x)
                   R(y)
         \left.\atop\right] = 0(h^3_{\mu\nu}),
\end{eqnarray}
where the metrics may be different at the points $x$ and $y$,
{\it i.e.} one may have $h^+_{\mu\nu}(x)$ and $h^-_{\mu\nu}(y)$,
respectively.

Let us now write the influence action $S^R_{IF}[h^{\pm}_{\mu\nu}]$
in a form which is more appropriate for the analysis of fluctuations
\cite{FVH}. We first decompose $S^R_{IF}$ into its real and imaginary
parts, because to the quadratic order in the metric perturbations the
real part gives information on the dissipation whereas the imaginary
part is related to noise,
\begin{equation}
   S^R_{IF}[h^{\pm}_{\mu\nu}]=\hat\Gamma_{IF}[h^{\pm}_{\mu\nu}]
                           +i\tilde\Gamma_{IF}[h^{\pm}_{\mu\nu}].
   \label{eq:real plus imaginary}
\end{equation}
For this we need to decompose the kernels
$\mbox{\rm K}^\pm(x-y;\bar\mu)$ and $\mbox{\rm K}(x-y)$ of
(\ref{eq:K}) into their real and imaginary parts also
\begin{eqnarray}
   \mbox{\rm K}^\pm(x-y;\bar\mu)
        &=&\hat\gamma_e(x-y;\bar\mu)\pm i\tilde\gamma_e(x-y),
           \nonumber \\
   \mbox{\rm K}(x-y)
        &=&\hat\gamma_o(x-y) - i\tilde\gamma_e(x-y),
   \label{eq:K kernels}
\end{eqnarray}
where
\begin{eqnarray}
   \hat\gamma_e(x-y;\bar\mu)
        &=&-{1\over 2}\int {d^4p\over(2\pi)^4}
                           \cos p\cdot (x-y)
                           \ln {|p^2|\over {\bar\mu}^2},
           \nonumber \\
   \tilde\gamma_e(x-y)
        &=& {\pi\over 2}\int {d^4p\over(2\pi)^4}
                             \cos p\cdot (x-y)
                             \theta(-p^2),
           \nonumber \\
   \hat\gamma_o(x-y)
        &=& {\pi\over 2}\int {d^4p\over(2\pi)^4}
                             \sin p\cdot (x-y)
                             \theta(-p^2)
                             sgn(-p^0).
   \label{eq:gamma kernels}
\end{eqnarray}
{}From the above definitions it is easy to show that these kernels
verify
the following symmetry relations:
\begin{equation}
   \hat\gamma_e(x-y;\bar\mu) = \hat\gamma_e(y-x;\bar\mu), \,\,\,\,\,
   \tilde\gamma_e(x-y)       = \tilde\gamma_e(y-x), \,\,\,\,\,
   \hat\gamma_o(x-y)         = -\hat\gamma_o(y-x).
\end{equation}
It is also convenient to introduce two new kernels
$\hat\gamma(x-y;\bar\mu)$ and $\mbox{\rm H}(x-y;\bar\mu)$ as
\begin{eqnarray}
   \hat\gamma(x-y;\bar\mu)
        &=& \hat\gamma_e(x-y;\bar\mu)-\hat\gamma_o(x-y)sgn(x^0-y^0)
           =\hat\gamma(y-x;\bar\mu),
           \nonumber \\
   \mbox{\rm H}(x-y;\bar\mu)
        &=& \hat\gamma_e(x-y;\bar\mu)+\hat\gamma_o(x-y).
\end{eqnarray}
With these definitions the real and imaginary parts of
(\ref{eq:real plus imaginary})
may be written, respectively, as
\begin{equation}
   \hat\Gamma_{IF}[h^{\pm}_{\mu\nu}]
        =\frac{3\alpha}{2}
           \int d^4xd^4y
               \left[ C^{+}_{\mu\nu\alpha\beta}(x)
                     -C^{-}_{\mu\nu\alpha\beta}(x)
               \right]\mbox{\rm H}(x-y;\bar\mu)
               \left[ C^{+\mu\nu\alpha\beta}(y)
                     +C^{-\mu\nu\alpha\beta}(y)
               \right],
   \label{eq:real}
\end{equation}
\begin{equation}
   \tilde\Gamma_{IF}[h^{\pm}_{\mu\nu}]
          =\frac{3\alpha}{2}
           \int d^4xd^4y
               \left[ C^{+}_{\mu\nu\alpha\beta}(x)
                     -C^{-}_{\mu\nu\alpha\beta}(x)
               \right] \tilde\gamma_e(x-y)
               \left[ C^{+\mu\nu\alpha\beta}(y)
                     -C^{-\mu\nu\alpha\beta}(y)
               \right],
   \label{eq:imaginary}
\end{equation}
Note that the integrand of the imaginary part depends (quadratically)
only on the differences of the field at the points $x$ and $y$ as one
expects of a term which is going to be the origin of noise. In
order to write the influence functional in the standard
Feynman-Vernon form \cite{FVH} it is
convenient to change the limits in the time integrations of the
above expressions so that $x^0 > y^0$ always in the dissipation and
fluctuation terms. For this we use the following further
symmetries of the above kernels
\begin{eqnarray}
           \hat\gamma_e(x^0-y^0,{\bf x}-{\bf y};\bar\mu)
        &=&\hat\gamma_e(y^0-x^0,{\bf x}-{\bf y};\bar\mu),
           \,\,\,\,\,
           \tilde\gamma_e(x^0-y^0,{\bf x}-{\bf y})
         = \tilde\gamma_e(y^0-x^0,{\bf x}-{\bf y}),
           \nonumber \\
           \hat\gamma(x^0-y^0,{\bf x}-{\bf y};\bar\mu)
        &=&\hat\gamma(y^0-x^0,{\bf x}-{\bf y};\bar\mu),
           \,\,\,\,\,\,\,
           \hat\gamma_o(x^0-y^0,{\bf x}-{\bf y})
         =-\hat\gamma_o(y^0-x^0,{\bf x}-{\bf y}),
           \nonumber \\
\end{eqnarray}
and that, for an arbitrary function $f(x^0,y^0)$, the integral
$\int_{-\infty}^{\infty}dx^o\int_{-\infty}^{\infty}dy^of(x^0,y^0)$ is
$2\int_{-\infty}^{\infty}dx^o\int_{-\infty}^{x^o}dy^of(x^0,y^0)$
if $f(x^0,y^0)=f(y^0,x^0)$ and is zero if
$f(x^0,y^0)=-f(y^0,x^0)$. Since the kernels $\hat\gamma_o(x-y)$
and $\tilde\gamma_e(x-y)$ will be related respectively to
dissipation and noise we will introduce the dissipation and noise
kernels as follows
\begin{eqnarray}
   \mbox{\rm D}(x-y)
        &=&-3\alpha\hat\gamma_o(x-y),
           \nonumber \\
   \mbox{\rm N}(x-y)
        &=&3\alpha\tilde\gamma_e(x-y).
   \label{eq:dissipation and noise kernels}
\end{eqnarray}
Finally we can write the influence functional in the standard form as
\begin{eqnarray}
   S^R_{IF}[h^{\pm}_{\mu\nu}]
     &=&\frac{3\alpha}{2}
        \int d^4x d^4y\hat\gamma(x-y;\bar\mu)
        \left[ C^{+}_{\mu\nu\alpha\beta}(x)
               C^{+\mu\nu\alpha\beta}(y)
              -C^{-}_{\mu\nu\alpha\beta}(x)
               C^{-\mu\nu\alpha\beta}(y)
        \right]
        \nonumber \\
     & &-\int_{-\infty}^{\infty}dx^o\int_{-\infty}^{x^o}dy^o
        \int d^3{\bf x} d^3{\bf y}
        \Delta C_{\mu\nu\alpha\beta}(x)
        \mbox{\rm D}(x-y)
        \{ C^{\mu\nu\alpha\beta}(y)\}
        \nonumber \\
     & &+i\int_{-\infty}^{\infty}dx^o\int_{-\infty}^{x^o}dy^o
        \int d^3{\bf x} d^3{\bf y}
        \Delta C_{\mu\nu\alpha\beta}(x)
        \mbox{\rm N}(x-y)
        \Delta C^{\mu\nu\alpha\beta}(y).
   \label{eq:real plus imaginary bis}
\end{eqnarray}
where we have used the notation
\begin{equation}
   \Delta C(x) \equiv  C^+(x)-C^-(x),
   \,\,\,\,\,\,\,\,\,
   \{ C(x)\} \equiv  C^+(x)+C^-(x),
\end{equation}
for the difference and sum of the $+$ and $-$ fields at the same
spacetime point.

The first term in the influence functional is non-local and will
contribute to give a non local stress tensor in the semiclassical
equations but does not mix the $+$ and $-$ fields. On the other
hand, in the second and third terms, which are also non local, there
is mixing of the $+$ and $-$ fields. The second will contribute
to the dissipation and the third to noise as we will see in the
next section. We should remark also the explicit gauge invariant form
of the influence functional above.


\subsection{The fluctuation-dissipation relation}
\label{subsec:FD relation}


It is easy to derive a relation between the dissipation and noise
kernels in our case in analogy with the quantum Brownian model
of ref. \cite{HPY93} or the case of a quantum scalar field in an
anisotropic Bianchi I cosmology \cite{HS95}.

We first note that the dissipation kernel $\mbox{\rm D}(x)$ of
(\ref{eq:dissipation and noise kernels}) can be written as a time
derivative:
\begin{equation}
   \mbox{\rm D}(x)=\frac{\partial}{\partial\eta}\gamma(x),
\end{equation}
where
\begin{equation}
   \gamma(x)={3\pi\alpha\over 2}
             \int {d^4p\over(2\pi)^4}
                  \frac{e^{ip\cdot x}}{|p^o|}
                  \theta(-p^2),
\end{equation}
where $1/|p^o|$ in this integral must be understood as its
Hadamard's finite part distribution.
Then the fluctuation-dissipation relation takes the form
\begin{equation}
   \mbox{\rm N}(x)
        =\int d^4x' \mbox{\rm K}_{FD}(x-x')\gamma(x'),
   \label{eq:FD relation}
\end{equation}
where the fluctuation-dissipation kernel $\mbox{\rm K}_{FD}(x-x')$
is given by the distribution

\begin{equation}
   \mbox{\rm K}_{FD}(x-x')=\delta^3({\bf x}-{\bf x}')
                           \int_{0}^{\infty}\frac{dq^o}{\pi}q^o
                           \cos{q^o (\eta-\eta')}.
   \label{eq:KFD}
\end{equation}
To check this fluctuation-dissipation relation one may simply
substitute (\ref{eq:KFD})
into the right hand side of (\ref{eq:FD relation}), perform the
integrations and the result is the noise kernel defined by
(\ref{eq:gamma kernels}) and
(\ref{eq:dissipation and noise kernels}). This relation is important
since it gives a direct connection between the effect of quantum
fluctuations of the environment on the dissipation of the
gravitational field inhomogeneities.


\section{Effective influence action and stochastic semiclassical
         equations}
\label{sec:field equations}


In this section we obtain from the influence action of Sec.
\ref{sec:influence action} the effective influence action. From
that effective action the explicit semiclassical
Einstein-Langevin equations will follow.

We start by recalling some well known relations of frequent use
in statistical physics. The path integral Gaussian identity
\cite{FVH},
$\int {\cal D}\xi \exp[-{1\over2}(\xi,L\xi)+(b,\xi)+c]
        = (\det L)^{-1/2}\exp[{1\over2}(b,L^{-1}b)-c]$,
where $L$ is a linear operator acting on the field $\xi(x)$ and
the brackets in the exponents stand for $(\xi,\zeta)=
\int \xi(x) \zeta(x)d^nx$, can be used to show that (change $L$
by $A^{-1}$, $b$ by $ik$ and take $c=0$)
\begin{equation}
   \Phi[k]\equiv e^{-{1\over2}(k,A k)}
        = \int {\cal D}\xi {\cal P}[\xi]e^{i(k,\xi)},
   \label{eq:characteristic functional}
\end{equation}
where ${\cal P}[\xi]$ is given by
\begin{equation}
   {\cal P}[\xi]=
        \frac{e^{-{1\over2}(\xi,A^{-1}\xi)}}
        {\int {\cal D}\xi e^{-{1\over2}(\xi,A^{-1}\xi)}}\,\,\,.
   \label{eq:distribution functional}
\end{equation}
If we interpret ${\cal P}[\xi]$ as the probability distribution
functional of the field $\xi(x)$ then $\Phi[k]$, which is its
functional Fourier transform, is the characteristic functional.
The mean value of a given functional ${\cal A}[\xi]$ is defined
by
\begin{equation}
   \langle {\cal A}\rangle_\xi
        = \int {\cal D}\xi{\cal P}[\xi]{\cal A}[\xi]
   \label{eq:mean value}
\end{equation}
and the $n$-correlation functions of the field $\xi(x)$ may be
derived from the characteristic functional as
\begin{equation}
   \langle \xi(x^1)\cdots\xi(x^n)\rangle_\xi
        =\left.\frac{1}{i^n}
               \frac{\delta^n\Phi[k]}
                    {\delta k(x^1)\cdots\delta k(x^n)}
         \right|_{k=0}.
   \label{eq:correlation functions}
\end{equation}
Since in this case the characteristic functional is Gaussian we have
that $\langle \xi(x)\rangle_\xi =0$ and that the two point
correlation function is $\langle \xi(x)\xi(x')\rangle_\xi =A(x,x')$.

Let us go back now to the influence functional derived in Sec.
\ref{sec:influence action},
\begin{equation}
   {\cal F}_{IF}[h^{\pm}_{\mu\nu}]
        \equiv e^{iS^R_{IF}[h^{\pm}_{\mu\nu}]}
        =      e^{i\hat\Gamma_{IF}[h^{\pm}_{\mu\nu}]}
               e^{-\tilde\Gamma_{IF}[h^{\pm}_{\mu\nu}]}
   \label{eq:influence functional}
\end{equation}
where the real and imaginary part of the influence action
$\hat\Gamma_{IF}$ and $\tilde\Gamma_{IF}$, respectively, are given
in (\ref{eq:real plus imaginary bis}). We recall now the
observation that the imaginary part of the influence action
depends on the difference between the $+$ and $-$ fields only,
more precisely, it can be written as
$\tilde\Gamma_{IF}[h^{\pm}_{\mu\nu}(x)]=
 \tilde\Gamma_{IF}[\Delta C^{\mu\nu\alpha\beta}(x)]$.
This is the signal
\cite{FVH} that the effect of the environment on the system,
given by this part of the influence functional, is equivalent
to a classical stochastic source $\xi_{\mu\nu\alpha\beta}(x)$
(a tensor field in this case) whose probability distribution has
such part of the influence functional as its characteristic
functional.
In fact, using (\ref{eq:characteristic functional}) we see that
(\ref{eq:influence functional}) can be written as
\begin{equation}
   {\cal F}_{IF}[h^{\pm}_{\mu\nu}]
   =\int {\cal D}\xi {\cal P}\left[\xi\right]
    \exp i
      \left\{\hat\Gamma_{IF}[h^{\pm}_{\mu\nu}]
            +\int d^4x \xi_{\mu\nu\alpha\beta}(x)
            \Delta C^{\mu\nu\alpha\beta}(x)
      \right\},
\end{equation}
where ${\cal P}[\xi]$ is the Gaussian probability
distribution given by (see (\ref{eq:distribution functional}) and
(\ref{eq:real plus imaginary bis}))
\begin{equation}
   {\cal P}\left[\xi\right]=
   \frac{e^{-\frac{1}{2}\int d^4xd^4y
         \xi(x)(\mbox{\scriptsize\rm N}(x-y))^{-1}\xi(y)}}
        {\int {\cal D}\xi
         e^{-\frac{1}{2}\int d^4xd^4y
         \xi(x)(\mbox{\scriptsize\rm N}(x-y))^{-1}\xi(y)}}.
   \label{eq:gaussian distribution}
\end{equation}
Therefore using (\ref{eq:mean value}) we can interprete the
influence functional as the following mean value
\begin{equation}
   {\cal F}_{IF}[h^{\pm}_{\mu\nu}] =
           \langle e^{iS^{eff}_{IF}[h^{\pm}_{\mu\nu},\xi]}
           \rangle_\xi
\end{equation}
where the effective influence action is defined by
\begin{equation}
   S^{eff}_{IF}[h^{\pm}_{\mu\nu},\xi]
        = \hat\Gamma_{IF}[h^{\pm}_{\mu\nu}]
         +\int d^4x \xi_{\mu\nu\alpha\beta}(x)
          \Delta C^{\mu\nu\alpha\beta}(x).
   \label{eq:S effective}
\end{equation}
The effect of the environment (quantum fields) on the system
(the gravitational field) is completely characterized by this
effective action, the tensor $\xi_{\mu\nu\alpha\beta}(x)$
plays the role of a stochastic source with the
Gaussian probability distribution given by
(\ref{eq:gaussian distribution}). This tensor has the symmetries
of the Weyl tensor, {\it i.e.} it has the
symmetries of the Riemann tensor and vanishing trace in all its
indices.
The kernel $\mbox{\rm N}$, which appears in
(\ref{eq:gaussian distribution}), can thus be interpreted as the
noise kernel in our problem. Since the probability distribution
is Gaussian the noise kernel is the two point correlation
function of the stochastic source. This source,
in fact, is completely characterized by the relations
\begin{eqnarray}
      \langle\xi_{\mu\nu\alpha\beta}(x)\rangle_{\xi}
  &=& 0,
      \nonumber \\
      \langle\xi_{\mu\nu\alpha\beta}(x)
      \xi_{\rho\sigma\lambda\theta}(y)\rangle_{\xi}
  &=& T_{\mu\nu\alpha\beta\rho\sigma\lambda\theta}
      \mbox{\rm N}(x-y),
   \label{eq:gaussian correlations}
\end{eqnarray}
where the explicit form of the tensor
$T_{\mu\nu\alpha\beta\rho\sigma\lambda\theta}$ is given in the
Appendix, it is the product of four metric tensors, in such
a combination that the right-hand side of the equation satisfy
the Weyl symmetries of the two stochastic fields on the
left-hand side.
It is easy to obtain these relations using
(\ref{eq:correlation functions}), note that
the characteristic functional has the form
$\Phi[k_{\mu\nu\alpha\beta}(x)]$, where $k_{\mu\nu\alpha\beta}(x)$
has the symmetries of the Weyl tensor.
It should be now clear also that
$\mbox{\rm D}$ is the dissipation kernel for this problem
since it is related to the noise by the fluctuation-dissipation
relation (\ref{eq:FD relation}). It is worth to mention that the
association of the imaginary terms of the effective action
(or the influence action) as terms coming from the interaction
of the field with stochastic sources has been used also in the
context of scalar fields in interaction with other fields, or
selfinteracting, in order to study the nonequilibrium dynamics
of these quantum fields \cite{gleiser}.


\subsection{Stochastic semiclassical equations}
\label{subsec:stochastic equations}


We are now in the position to derive the semiclassical equations
for the gravitational field due to a quantum scalar field.
We recall that the effect of the quantum
field on the gravitational field is given
by the effective influence action (\ref{eq:S effective}). Therefore
the total effective action, which includes the action of the
gravitational field plus the previous effective influence action is
given by
\begin{equation}
   S_{eff}[{\tilde g}^\pm,\xi]
        = S^R_g[\tilde g^+_{\mu\nu}]
         -S^R_g[\tilde g^-_{\mu\nu}]
         +S^{eff}_{IF}[h^{\pm}_{\mu\nu},\xi]
\end{equation}
where $S^R_g[\tilde g_{\mu\nu}]$ is the renormalized
action of the gravitational
field (\ref{eq:renormalized gravitational action}). Since this is
an effective action for the metric perturbations the field
equations for the metric can be derived in a similar way as
(\ref{eq:definition of field equations bis}). It is usually
convenient to introduce new variables
$
      \bar h_{\mu\nu}
   \equiv\left(h^+_{\mu\nu}+h^-_{\mu\nu}\right)/2,
      \Delta h_{\mu\nu}
   \equiv h^+_{\mu\nu}-h^-_{\mu\nu},
$
the average field and the difference field respectively. Then the
field equations are
$\left. \frac{\delta}{\delta\Delta h_{\mu\nu}}
       \left( S_{eff}\left[\omega,\bar h_{\mu\nu},
                              \Delta h_{\mu\nu},\xi
                        \right]
       \right)
 \right|_{\Delta h_{\mu\nu}=0}=0$; or in an equivalent form,
which is of more practical use to us,
\begin{equation}
   \left. \frac{\delta}{\delta h^+_{\mu\nu}}
          \left(S^R_g[\tilde g^+_{\mu\nu}]
                +\hat\Gamma_{IF}[h^\pm_{\mu\nu}]
                +\int d^4x \xi_{\mu\nu\alpha\beta}(x)
                C^{+\mu\nu\alpha\beta}(x)
          \right)
   \right|_{h^+_{\mu\nu}=h^-_{\mu\nu}}=0.
   \label{eq:semiclassical equations}
\end{equation}
The functional derivations needed in the first two terms can all
be found in Appendix E of Paper I. For the new term, we can
easily prove that
\begin{equation}
   \int d^4x \xi^{\alpha\sigma\tau\rho}(x)
             C^+_{\alpha\sigma\tau\rho}(x)
        =
   -2\int d^4x
        \partial_\sigma\partial_\rho\xi^{\alpha\sigma\tau\rho}(x)
        h^+_{\alpha\tau}(x),
     \label{eq:functional differentiation}
\end{equation}
where in this expression we have assumed that the tensor
field $\xi_{\mu\nu\alpha\beta}$ has the symmetries of
the Weyl tensor.

Finally the semiclassical equation (\ref{eq:semiclassical equations})
can be written as
\begin{eqnarray}
   & &
      e^{6\omega}\left[-{1\over 16\pi G_N}
                          \left( \tilde G^{\mu\nu}_{(0)}
                                +\tilde G^{\mu\nu}_{(1)}
                          \right)
                       -{\alpha\over 12}
                          \left( \tilde B^{\mu\nu}_{(0)}
                                +\tilde B^{\mu\nu}_{(1)}
                          \right)
                       +{\alpha\over 2}
                          \left( \tilde H^{\mu\nu}_{(0)}
                                +\tilde H^{\mu\nu}_{(1)}
                          \right)
                       -\alpha\tilde R^{(0)}_{\alpha\beta}
                              \tilde C^{\mu\alpha\nu\beta}_{(1)}
                 \right]
      \nonumber \\
   & &\hskip1cm
      +{3\alpha\over 2}
        \left[-4(C^{\mu\alpha\nu\beta}_{(1)}\omega )_{,\alpha\beta}
              +\int d^4y A^{\mu\nu}_{(1)}(y)\mbox{\rm H}(x-y;\bar\mu)
        \right]
      +F^{\mu\nu}[\xi]
      = O(h^2_{\mu\nu}),
   \label{eq:semiclassical equations bis}
\end{eqnarray}
where the $(0)$ and $(1)$ subindices mean the zero and one orders,
respectively, in terms of the perturbation $h_{\mu\nu}$. Terms
with and without tilde refer to tensors obtained with metrics
$\tilde g_{\mu\nu}$ and $g_{\mu\nu}$, respectively. $G^{\mu\nu}(x)$
is the Einstein tensor, $B^{\mu\nu}(x)$, $A^{\mu\nu}(x)$ and
$H^{\mu\nu}(x)$ are given in Paper I as:
\begin{eqnarray}
   B^{\mu\nu}(x)
      &\equiv& {1\over2}g^{\mu\nu}R^2
               -2RR^{\mu\nu}
               +2R^{;\mu\nu}
               -2g^{\mu\nu}\Box_g R,
         \nonumber \\
   A^{\mu\nu}(x)
      &\equiv& {1\over2}g^{\mu\nu}C_{\alpha\beta\rho\sigma}
                                  C^{\alpha\beta\rho\sigma}
               -2R^{\mu\alpha\beta\rho}{R^\nu}_{\alpha\beta\rho}
               +4R^{\mu\alpha}{R_\alpha}^\nu
         \nonumber \\
      & &\hskip .1cm
               -{2\over3}RR^{\mu\nu}
               -2\Box_g R^{\mu\nu}
               +{2\over3}R^{;\mu\nu}
               +{1\over3}g^{\mu\nu}\Box_g R,
         \nonumber \\
   H^{\mu\nu}(x)
      &\equiv& -R^{\mu\alpha}{R_\alpha}^\nu
               +{2\over3}RR^{\mu\nu}
               +{1\over2}g^{\mu\nu}R_{\alpha\beta}R^{\alpha\beta}
               -{1\over4}g^{\mu\nu}R^2.
\end{eqnarray}
The tensor $F^{\mu\nu}(x)$ results from the functional
variation of the stochastic term, it can be read off directly from
(\ref{eq:functional differentiation}),
\begin{equation}
   F^{\mu\nu}(x)=
       -2\partial_\alpha\partial_\beta\xi^{\mu\alpha\nu\beta}(x),
   \label{eq:source}
\end{equation}
and it is symmetric and traceless, {\it i.e.}
$F^{\mu\nu}(x)=F^{\nu\mu}(x)$ and $F^\mu_{\,\,\mu}(x)=0$, which means
that there is no stochastic correction to the trace anomaly.

{}From equations (\ref{eq:semiclassical equations bis}) we may now
define the effective stress tensor $T^{\mu\nu}_{eff}$, which is the
term to write on the right hand side of Einstein equations,
{\it i.e.} let us write (\ref{eq:semiclassical equations bis}) as
\begin{eqnarray}
   \tilde G^{\mu\nu}(x)
        &=& 8\pi G_N \left( T^{\mu\nu}_c
                           +T^{\mu\nu}_{eff}
                     \right)
           \nonumber \\
   T^{\mu\nu}_{eff}
        &\equiv& \langle T^{\mu\nu} \rangle_q
                +2e^{-6\omega}F^{\mu\nu}[\xi]
   \label{eq:effective stress tensor}
\end{eqnarray}
where $\langle T^{\mu\nu} \rangle_q$ is the (quantum) vacuum
expectation value of the stress tensor of the quantum field
up to first order in $h_{\mu\nu}$
and we have added a classical stress tensor $T^{\mu\nu}_c$ to
include the case in which there is also a classical source (this was
not considered from the beginning for simplicity). The
quantum stress tensor is given by
\begin{eqnarray}
   \langle T^{\mu\nu}_{(0)} \rangle_q
        &=& \alpha\left[ \tilde H^{\mu\nu}_{(0)}
                        -{1\over 6} \tilde B^{\mu\nu}_{(0)}
                  \right]
           \nonumber \\
   \langle T^{\mu\nu}_{(1)} \rangle_q
        &=& \alpha\left[ \atop \right.
                         (\tilde H^{\mu\nu}_{(1)}
                          -2\tilde R^{(0)}_{\alpha\beta}
                          \tilde C^{\mu\alpha\nu\beta}_{(1)}
                         )
                        -{1\over 6} \tilde B^{\mu\nu}_{(1)}
              \nonumber \\
          & &\hskip1cm
                        +3e^{-6\omega}
                            \left( -4( C^{\mu\alpha\nu\beta}_{(1)}
                                       \omega
                                     )_{,\alpha\beta}
                        +\int d^4y A^{\mu\nu}_{(1)}(y)
                                      \mbox{\rm H}(x-y;\bar\mu)
                           \right)
                  \left. \atop \right].
   \label{eq:quantum stress tensor}
\end{eqnarray}
This tensor was already given in Paper I and was first computed
by other means in Refs. \cite{HWS}. Now we have derived a stochastic
correction to this tensor which accounts for the noise due to the
fluctuations of the quantum field.

If we now take the mean value of equation
(\ref{eq:effective stress tensor}) with respect to the stochastic
source $\xi$ we find that, as a
consequence of (\ref{eq:gaussian correlations}),
\begin{equation}
          \langle T^{\mu\nu}_{eff}
          \rangle_\xi
        = \langle T^{\mu\nu} \rangle_q
\end{equation}
and we recover the semiclassical Einstein equations of Paper I.

The stochastic correction to the stress tensor has vanishing divergence
to the first order in the metric perturbations. In fact,
using that ${\tilde g}_{\mu\nu}=e^{2\omega}g_{\mu\nu}$ and that
$F^{\mu\nu}$ is symmetric and traceless, it is easy to see that
${\tilde\nabla}_\nu\left(e^{-6\omega}F^{\mu\nu}\right)=
e^{-6\omega}\nabla_\nu F^{\mu\nu}$, then from (\ref{eq:source})
and the symmetries of $\xi$ we obtain that
$\nabla_\nu F^{\mu\nu}=O(h_{\mu\nu})$. It is thus consistent to write
this term on the right-hand side of Einstein equations and consider it
as a correction of order higher than
$\langle T^{\mu\nu}_{(0)}\rangle_q$ (note that
${\tilde\nabla}_\nu \langle T^{\mu\nu}_{(0)}\rangle_q=O(h^2_{\mu\nu})$).

To summarize,  the stochastic
semiclassical equations (\ref{eq:effective stress tensor}) can be
called the semiclassical Einstein-Langevin equations for weakly
inhomogeneous spatially flat cosmologies in the presence of  a
conformally coupled massless scalar field. The usual semiclassical
equations can be seen as the mean value of these equations with
respect to a tensor field source $\xi$ with
Gaussian probability distributions
(\ref{eq:gaussian distribution}). This stochastic source couples
to the Weyl (conformal) tensor of the spacetime metric in the form
given by (\ref{eq:S effective}), this means that it has the symmetries
of that tensor and thus that it has only $10$ independent components
at each spacetime point.

The fact that the stochastic source couples to the conformal
tensor should not come as a surprise since we expect that, for a
conformal quantum field, non trivial quantum effects should be
a consequence of breaking the conformal symmetry of the spacetime,
which is characterized by the conformal tensor. For instance, it is
known that the probability density of pair creation in this case,
or in the presence of only small anisotropies, is determined by the
square of the Weyl tensor \cite{particle production}. Thus as it has
been emphasized in \cite{CH94} there is a direct relation between
particle creation and noise.


\section{Metric fluctuations in flat spacetime}
\label{sec:correlations}


In this section we make a simple application of the semiclassical
Einstein-Langevin equations obtained in the previous section to the
case in which the background spacetime is not cosmological,
{\it i.e.} when $\omega =0$ in (\ref{eq:FRWL}). This restriction
simplifies considerably the semiclassical equations. We will take
here a perturbative approach in which the semiclassical corrections
to Einstein's equations are seen as analytic perturbations
(in $\hbar$) to the classical Einstein's equations; see Simon
\cite{S91} for a justification of this point of view.

Let us assume that we have some weak stress tensor source
$T^c_{\mu\nu}$ in flat spacetime. For instance, we could have a
cosmic string or a bulk of Newtonian matter. Such a source will
produce some classical linear inhomogeneities $h^c_{\mu\nu}$ and the
spacetime metric will be
\begin{equation}
   g_{\mu\nu}=\eta_{\mu\nu}+h^c_{\mu\nu}.
\end{equation}
If we have a massless conformally coupled quantum scalar field in
this background, the stress tensor acting on the spacetime will now
be $T^c_{\mu\nu}+T^{eff}_{\mu\nu}$. The semiclassical
equations up to first order in the metric perturbations can be written
as
\begin{equation}
   G_{\mu\nu}(x)
        = 8\pi G_N \left[ T^c_{\mu\nu}(x)
                         +T^{eff}_{\mu\nu}(x)
                   \right],
   \label{eq:flat semiclassical equation}
\end{equation}
where $T^{eff}_{\mu\nu}$ follows from
(\ref{eq:effective stress tensor}) and (\ref{eq:quantum stress tensor})
with $\omega =0$; note that in this case
$\langle T^{\mu\nu}_{(0)}\rangle_q=0$.

The stochastic term $2F^{\mu\nu}$ will produce a stochastic
contribution $h^{st}_{\mu\nu}$ to the spacetime inhomogeneity,
{\it i.e.} $h_{\mu\nu}=h^c_{\mu\nu}+h^{st}_{\mu\nu}$. Let us now
consider such contribution.

Substituting into (\ref{eq:flat semiclassical equation}) and
taking into account that $h^c_{\mu\nu}$
already satisfies the classical equation, we obtain a linear
equation for the stochastic term $h^{st}_{\mu\nu}$,
which we may write as
\begin{eqnarray}
   \Box h^{st}_{\mu\nu}
        &=& 16\pi G_N S^{st}_{\mu\nu},
           \nonumber \\
   S^{\mu\nu}_{st}
        &=& 2F^{\mu\nu}
         =-4\partial_\alpha\partial_\beta\xi^{\mu\alpha\nu\beta},
\end{eqnarray}
together with the harmonic gauge condition,
$(h^{st}_{\mu\nu} - {1\over2}\eta_{\mu\nu}h^{st})^{,\nu}=0$, which
was used to write the previous equation.
The solution of these equations as a Cauchy problem with
boundary conditions
$h^{st}_{\mu\nu}(-\infty,{\bf x})
      = \partial_th^{st}_{\mu\nu}(-\infty,{\bf x})
      = 0$, is given by
\begin{equation}
   h^{st}_{\mu\nu}(t,{\bf x})
        = 16\pi G_N \int^{t}_{-\infty}dt'\int_{R^3}d^3{\bf x}
          D_R(t-t',{\bf x}-{\bf x}')S^{st}_{\mu\nu}(t',{\bf x}'),
   \label{eq:flat solution}
\end{equation}
where $D_R$ is the retarded Green's function
\begin{equation}
   D_R(x-x')=-\frac{1}{4\pi |{\bf x}-{\bf x}'|}
              \delta(t-t'-|{\bf x}-{\bf x}'|).
\end{equation}
Note that after imposing the above boundary conditions for
$h^{st}_{\mu\nu}$, the gauge has been completely fixed because
two metric components within the harmonic gauge can differ
by $2\zeta_{\mu,\nu}$ where $\zeta_\mu$ is an harmonic vector
field but this vector field is zero when such boundary conditions
are imposed.
Since $S^{st}_{\mu\nu}$ is linear in the stochastic source, it is
obvious from (\ref{eq:flat solution}) that
$\langle h^{st}_{\mu\nu}(x) \rangle_\xi=0$.

Let us now compute the two point correlation function of the
stochastic metric fluctuations $h^{st}_{\mu\nu}$
\begin{equation}
   \langle h^{st}_{\mu\nu}(t,{\bf x})
           h^{st}_{\lambda\theta}(s,{\bf y})
   \rangle_\xi
     =(4G_N)^2\int_{R^3}d^3{\bf x}'\int_{R^3}d^3{\bf y}'
      \frac{\langle
             S^{st}_{\mu\nu}(t-|{\bf x}-{\bf x}'|,{\bf x}')
             S^{st}_{\lambda\theta}(s-|{\bf y}-{\bf y}'|,{\bf y}')
            \rangle_\xi
           }
           {|{\bf x}-{\bf x}'||{\bf y}-{\bf y}'|
           },
   \label{eq:correlation for the flat solution}
\end{equation}
here we made use of (\ref{eq:flat solution}) and performed two
time integrations. Now since $S^{st}_{\mu\nu}$ has the form of
a linear operator acting on the tensorial  stochastic
source, $\xi$, the correlation function of
$S^{st}_{\mu\nu}$ can be written in terms of the correlation
function of that source, {\it i.e.} in terms of the noise
kernel $\mbox{\rm N}$. After some simple manipulations
we obtain
\begin{equation}
   \langle
        S^{\mu\nu}_{st}(x)S^{\lambda\theta}_{st}(y)
   \rangle_\xi
        = {3\pi\alpha\over2}\Theta^{\mu\nu\lambda\theta}
          \int {d^4p\over(2\pi)^4} e^{ip\cdot\sigma}
          \theta(-p^2),
\end{equation}
where we have defined $\sigma^\mu\equiv x^\mu - y^\mu$, and
$\Theta^{\mu\nu\lambda\theta}$ is the operator
\begin{equation}
\Theta^{\mu\nu\lambda\theta}
        \equiv \frac{2}{3}
          \left[ 3\hat P^{\lambda(\mu}\hat P^{\nu)\theta}
                - \hat P^{\mu\nu}\hat P^{\lambda\theta}
          \right],
\end{equation}
with $\hat P^{\mu\nu}=\eta^{\mu\nu}\Box - \partial^\mu\partial^\nu$,
where all derivatives are with respect to $\sigma^\mu$.
Substituting this
into (\ref{eq:correlation for the flat solution}) leads to
\begin{equation}
   \langle h^{st}_{\mu\nu}(t,{\bf x})
           h^{st}_{\lambda\theta}(s,{\bf y})
   \rangle_\xi
        = 6\pi^7\alpha\left(32G_N\right)^2
          \Theta_{\mu\nu\lambda\theta}
          \int {d^4p\over(2\pi)^4} \theta(-p^2) e^{ip\cdot\sigma}
          \Delta^+(p)\Delta^-(p)
\end{equation}
where $\Delta^\pm$ are the Wightman functions. Finally,
after some further manipulations the correlation function of
$h^{st}_{\mu\nu}(x)$ can be written as:
\begin{equation}
   \langle h^{st}_{\mu\nu}(t,{\bf x})
           h^{st}_{\lambda\theta}(s,{\bf y})
   \rangle_\xi
        =(16\pi G_N)^2
         \Theta_{\mu\nu\lambda\theta}
         \int d^4z \mbox{\rm K}_{reg}(\sigma - z)
         \mbox{\rm N}(z),
   \label{eq:por fin}
\end{equation}
where $\mbox{\rm K}_{reg}(x)$ is the appropriately regularized
kernel defined by the Fourier transform integral
$\int d^4p \exp (ip\cdot x)\Delta^+(p)\Delta^-(p)\equiv
 \mbox{\rm K}_{reg}(x)$. Since the noise kernel is non-local,
see equations (\ref{eq:gamma kernels}) and
(\ref{eq:dissipation and noise kernels}), the noise is colored. A
simple order of magnitude estimate of the above result gives
that it is of the
order of $(\mbox{\rm Planck time}/\mbox{\rm time interval})^4$, as
one should expect of quantum fluctuations. Thus if we measure the
gravitational potential at a given point at different time intervals
we should find variations in this field of the order
$\sqrt{\langle h^{st}_{00}h^{st}_{00} \rangle}\sim
 (\mbox{\rm Planck time}/\mbox{\rm time interval})^2$.


\appendix



\section{}


The $T_{\mu\nu\alpha\beta\rho\sigma\lambda\theta}$ tensor
is given by
\begin{eqnarray}
  T_{\mu\nu\alpha\beta\rho\sigma\lambda\theta}
    &\equiv & \frac{1}{24}
          \left\{ 8\left[ \eta_{\rho[\mu}\eta_{\nu]\sigma}
                          \eta_{\lambda[\alpha}\eta_{\beta]\theta}
                         +\eta_{\rho[\alpha}\eta_{\beta]\sigma}
                          \eta_{\lambda[\mu}\eta_{\nu]\theta}
                         +\eta_{\alpha[\mu}\eta_{\nu]\beta}
                          \eta_{\lambda[\rho}\eta_{\sigma]\theta}
                   \right]\right.
          \nonumber \\
     & &  \hskip .5cm
                 +4\left[ \eta_{\rho[\mu}\eta_{\beta]\sigma}
                          \eta_{\lambda[\alpha}\eta_{\nu]\theta}
                         +\eta_{\rho[\mu}\eta_{\alpha]\sigma}
                          \eta_{\lambda[\nu}\eta_{\beta]\theta}
                         +\eta_{\rho[\nu}\eta_{\alpha]\sigma}
                          \eta_{\lambda[\beta}\eta_{\mu]\theta}
                         +\eta_{\rho[\beta}\eta_{\nu]\sigma}
                          \eta_{\lambda[\alpha}\eta_{\mu]\theta}
                   \right]
          \nonumber \\
     & &  \hskip .5cm
                 -3\left[\eta_{\mu\alpha}
                      \left( \eta_{\rho\lambda}\eta_{\sigma(\nu}
                             \eta_{\beta)\theta}
                            +\eta_{\sigma\theta}\eta_{\rho(\nu}
                             \eta_{\beta)\lambda}
                            -\eta_{\sigma\lambda}\eta_{\rho(\nu}
                             \eta_{\beta)\theta}
                            -\eta_{\rho\theta}\eta_{\sigma(\nu}
                             \eta_{\beta)\lambda}
                      \right)
                   \right.
          \nonumber \\
     & &  \hskip 1cm
                        +\eta_{\nu\beta}
                      \left( \eta_{\rho\lambda}\eta_{\sigma(\mu}
                             \eta_{\alpha)\theta}
                            +\eta_{\sigma\theta}\eta_{\rho(\mu}
                             \eta_{\alpha)\lambda}
                            -\eta_{\sigma\lambda}\eta_{\rho(\mu}
                             \eta_{\alpha)\theta}
                            -\eta_{\rho\theta}\eta_{\sigma(\mu}
                             \eta_{\alpha)\lambda}
                      \right)
          \nonumber \\
     & &  \hskip 1cm
                        -\eta_{\nu\alpha}
                      \left( \eta_{\rho\lambda}\eta_{\sigma(\mu}
                             \eta_{\beta)\theta}
                            +\eta_{\sigma\theta}\eta_{\rho(\mu}
                             \eta_{\beta)\lambda}
                            -\eta_{\sigma\lambda}\eta_{\rho(\mu}
                             \eta_{\beta)\theta}
                            -\eta_{\rho\theta}\eta_{\sigma(\mu}
                             \eta_{\beta)\lambda}
                      \right)
          \nonumber \\
     & &  \hskip 1cm
          \left.
                   \left.-\eta_{\mu\beta}
                      \left( \eta_{\rho\lambda}\eta_{\sigma(\nu}
                             \eta_{\alpha)\theta}
                            +\eta_{\sigma\theta}\eta_{\rho(\nu}
                             \eta_{\alpha)\lambda}
                            -\eta_{\sigma\lambda}\eta_{\rho(\nu}
                             \eta_{\alpha)\theta}
                            -\eta_{\rho\theta}\eta_{\sigma(\nu}
                             \eta_{\alpha)\lambda}
                      \right)
                   \right]
          \right\},
\end{eqnarray}
it comes from the functional derivative of the Weyl tensor with
respect to itself taking into account all its symmetries.


\acknowledgments


We are grateful to Esteban Calzetta and Bei-Lok Hu for very helpful
suggestions and discussions, and Rosario Mart{\'{\i}}n for a critical
reading of the manuscript. This work has been
partially supported by the CICYT Research Project number
\mbox{AEN93-0474}, and the European Project number
\mbox{CI1-CT94-0004}.




\begin{thebibliography}{11}




\bibitem{WFK}
    R. M. Wald,
          {\sl Commun. Math. Phys.} {\bf 54}, 1 (1977);
    L. H. Ford,
          {\sl Ann. Phys.} {\bf 144}, 238 (1982);
    L. H. Ford and C.-I Kuo,
          {\sl Phys. Rev.} {\bf D47}, 4510 (1993).

\bibitem{H89}
    B.-L. Hu,
          {\sl Physica} {\bf A158}, 399 (1989).

\bibitem{CH94}
    E. Calzetta and B.-L. Hu,
          {\sl Phys. Rev.} {\bf D49}, 6636 (1994).

\bibitem{HS95}
    B.-L. Hu and S. Sinha,
          {\sl Phys. Rev.} {\bf D51}, 1587 (1995).

\bibitem{HM95}
    B.-L. Hu and A. Matacz,
          {\sl Phys. Rev.} {\bf D51}, 1577 (1995).

\bibitem{FVH}
    R. P. Feynman and F. L. Vernon,
          {\sl Ann. Phys.} {\bf 24}, 118 (1963);
    R. P. Feynman and A. R. Hibbs,
          {\sl Quantum mechanics and path integrals},
               McGraw-Hill, New York (1965).

\bibitem{SKC}
    J. Schwinger,
          {\sl J. Math. Phys.} {\bf 2}, 407 (1961);
    L. V. Keldysh,
          {\sl Zh. Eksp. Teor. Fiz.} {\bf 47}, 1515 (1964)
          [{\sl Sov. Phys. JETP} {\bf 20}, 1018 (1965)];
    K. Chou, Z. Su, B. Hao, and L. Yu,
          {\sl Phys. Rep.} {\bf 118}, 1 (1985).

\bibitem{J86}
    R. D. Jordan,
          {\sl Phys. Rev.} {\bf D33}, 444 (1986).

\bibitem{CH87}
    E. Calzetta and B.-L. Hu,
          {\sl Phys. Rev.} {\bf D35}, 495 (1987).

\bibitem{CH89}
    E. Calzetta and B.-L. Hu,
          {\sl Phys. Rev.} {\bf D40}, 656 (1989).

\bibitem{CV94}
    A. Campos and E. Verdaguer,
          {\sl Phys. Rev.} {\bf D49}, 1861 (1994), Paper I.

\bibitem{HWS}
    G. T. Horowitz and R. M. Wald,
          {\sl Phys. Rev.} {\bf D21}, 1462 (1980);
          {\bf D25}, 3408 (1982);
    A. A. Starobinsky,
          {\sl Pis'ma Zh. Eksp. Teor. Fiz.} {\bf 34}, 460 (1981)
          [{\sl JETP Lett.} {\bf 34}, 438 (1981).

\bibitem{P90}
    J. P. Paz,
          {\sl Phys. Rev.} {\bf D42}, 529 (1990);
    D. Boyanovsky, H. J. de Vega, R. Holman, D.-S. Lee and A. Singh,
          {\sl Phys. Rev.} {\bf D51}, 4419 (1995).

\bibitem{CH95}
    E. Calzetta and B.-L. Hu,
          {\sl ``Quantum fluctuations, decoherence of the mean
           field, and structure formation in the early universe},
           gr-qc/9505046 (1995).

\bibitem{P90(2)}
    J. P. Paz,
          {\sl Phys. Rev.} {\bf D41}, 1054 (1990).

\bibitem{AL73}
    E. S. Abers and B. W. Lee,
          {\sl Phys. Rep.} {\bf 9C}, 1 (1973);
    P. Ramond,
          {\sl Field theory: a moder primer},
               Benjaming Cummings, Reading, Massachusetts (1965).

\bibitem{L75}
    G. Leibbrandt,
          {\sl Rev. Mod. Phys.} {\bf 47}, 849 (1975).

\bibitem{HPY93}
    B.-L. Hu, J. P. Paz and Y. Yang,
          {\sl Phys. Rev.} {\bf D47}, 1576 (1973).

\bibitem{gleiser}
    M. Morikawa,
          {\sl Phys. Rev.} {\bf D33}, 3607 (1986);
    D.-S. Lee and D. Boyanovsky,
          {\sl Nucl. Phys.} {\bf B406}, 631 (1993);
    M. Gleiser and R. O. Ramos,
          {\sl Phys. Rev.} {\bf D50}, 2441 (1994).

\bibitem{particle production}
    N. D. Birrell and P. C. W. Davies,
          {\sl J. Phys.} {\bf A13}, 2109 (1980);
    J. B. Hartle and B.-L. Hu,
          {\sl Phys. Rev.} {\bf D21}, 2756 (1980);
    J. A. Frieman,
          {\sl Phys. Rev.} {\bf D39}, 389 (1989);
    J. C{\'espedes} and E. Verdaguer,
          {\sl Phys. Rev.} {\bf D41}, 1020 (1990);
    A. Campos and E. Verdaguer,
          {\sl Phys. Rev.} {\bf D45}, 4428 (1992).

\bibitem{S91}
    J. Z. Simon,
          {\sl Phys. Rev.} {\bf D43}, 3308 (1991).




\end{thebibliography}
\end{document}